\documentclass[12pt]{article}
\newif\ifpdf
\ifx\pdfoutput\undefined
\pdffalse 
\else
\pdfoutput=1 
\pdftrue
\fi

\ifpdf
    \pdfcatalog{ /PageMode (/UseNone)
                 /OpenAction (fitbh)
                }
   \usepackage[pdftex,pdftitle={A THEORY OF ASYMMETRIC RESONANT
     CAVITIES},
urlcolor=blue,colorlinks=true]{hyperref}
   \usepackage[pdftex]{graphicx}
\else
   \usepackage[ps2pdf,pdftitle={PDF},urlcolor=blue,colorlinks=true]{hyperref}
   \usepackage{graphicx}
\fi
\include{cite.sty}
\newcommand{\nin}{\noindent}

\newcommand{\be}{\begin{equation}}
\newcommand{\ee}{\end{equation}}
\newcommand{\bea}{\begin{eqnarray}}
\newcommand{\eea}{\end{eqnarray}}

\newcommand{\sinchi}{\sin \chi}

\begin{document}
\ifpdf
\DeclareGraphicsExtensions{.pdf, .eps, .tif}
\else
\DeclareGraphicsExtensions{.eps}
\fi
\begin{center}
\vspace{15mm}

{\bf CHAOTIC LIGHT: A THEORY OF ASYMMETRIC RESONANT CAVITIES}
\vspace{20mm}

{JENS U.~N{\"O}CKEL\footnote{\href{http://darkwing.uoregon.edu/~noeckel}{\em Current address}: Department of Physics,
    University of Oregon, 1371 E 13th Avenue, Eugene, OR 97403}
AND A.~DOUGLAS STONE\\}
{\footnotesize{\it Department of Applied Physics and Physics, Yale University
\\P.~O.~Box 8284, New Haven Connecticut 06520-8284
\footnote{
{\em Published in} Optical Processes in Microcavities, edited by
R.K.Chang and A.J.Campillo (World Scientific Publishers, 1996)}
}}
\vspace{10mm}

{ABSTRACT}\\
\parbox[t]{30pc}{
\nin
{\footnotesize Spherical and cylindrical dielectric cavities support high Q
whispering gallery modes due to total internal reflection of the
trapped light.  When such a cavity is deformed smoothly the ray
dynamics of these modes becomes chaotic in a manner determined by the
KAM theory of classical hamiltonian dynamics.  The universal
properties of the ray dynamics predicted by KAM theory allow a general
understanding of the whispering gallery modes of such asymmetric
resonant cavities (ARCs).  This theory combined with simulations of
the non-linear map describing the ray motion provides the basis for a
ray-optics model of the Q-spoiling of these whispering gallery modes
for large deformations (greater than 1\% of the radius).  The model
predicts a sharp onset as a function of deformation for significant
Q-spoiling of these modes and highly directional emission above this
threshold.  Solution of the wave equation for typical whispering
gallery modes confirms the qualitative behavior predicted by the
ray-optics model even when the cavity is only a few times the resonant
wavelength. The model explains for the first time the lasing intensity
profile of highly deformed lasing droplets.}
}

\end{center}

\newpage

\section{Introduction}
\subsection{The Asymmetric Resonant Cavity (ARC)}
Optical processes in microcavities are currently under intensive study
\cite{Yamamoto} with several goals: to design novel microlasers
\cite{Slusher1,Slusher2,QELS} and other optical
devices \cite{Arnold}, to study non-linear optical effects
\cite{dropreview}, and to investigate the
interaction of atoms with single cavity modes \cite{Haroche}.  
The cavities employed
in these investigations can be divided into two categories:
Fabry-Perot type cavities of various designs, and dielectric cavities
(spheres or cylinders) in which the surface of the dielectric confines
certain modes of the electromagnetic field by total internal
reflection.  In the dielectric cavities these high-Q modes are the
``whispering gallery'' (WG) modes in which light rays circulate just
inside dielectric surface, propagating almost tangent to it
\cite{WG}.  The research described below pertains exclusively to the whispering gallery modes
of dielectric cavities.  For obvious geometric reasons such high-Q
modes are only possible when at least one cross-section of the cavity
is bounded by a convex curve (i.e. a curve which has positive
curvature everywhere) as is the case for a cylinder and a sphere.
However we are not primarily concerned here with the WG modes of ideal
dielectric spheres or cylinders but instead with the WG modes of a new
class of resonant cavities which are created by deformations of these
ideal symmetric dielectric cavities. The deformations considered can
be quite large, ranging from 1-50\% of the undeformed radius; the
crucial assumption is that they maintain the convexity of the
cavities.  While the WG modes of a spherical or cylindrical cavity can
be treated analytically and the effect of small deformations included
by perturbation theory \cite{young}, the deformations we consider are
so large that the modes of the deformed cavities are not
perturbatively related to those of the ideal cavity.  Hence we shall
refer to the class of deformed cavities we study as {\it asymmetric
resonant cavities} (ARCs) since their properties are not derivable
from those of the symmetric cavities (in any simple manner).  Because
the modes of ARCs are not simply related to those of symmetric
cavities it is not immediately obvious that high-Q WG modes still
exist in such structures; however we will show below that general
principles of non-linear dynamics imply that under rather general
conditions such modes exist if the cavity surface remains convex.

\subsection{ARCs and Quantum Chaos}
Because ARCs strongly violate at least one spatial symmetry of the
undeformed cavity, the wave equation describing their modes or
resonances is not separable in at least two of the three spatial
dimensions and the solutions are not any of the familiar special
functions of mathematical physics \cite{ellipse}.  Instead the wave
equation can only be solved numerically, e.g. by the T-matrix method
\cite{tmatrix} or by wavefunction matching techniques \cite{wilton}
of the type we
use in subsection 7.2 below.  It is for this reason that one does
not find the properties of such cavities described in advanced optics
textbooks or even in the research literature.  Furthermore, brute
force solution of the wave equation for a few modes in a given
geometry usually does not yield much physical insight into the general
properties of ARCs.  However the properties of wave-equations with
reduced or zero spatial symmetry have been extensively studied during
the past decade in another context, the attempt to understand the
quantum mechanics of systems which have either partially or fully
chaotic classical dynamics.  This field of inquiry is widely referred
to by the unfortunate descriptor ``quantum chaos'', although it is now
generally agreed that there exists no quantum analogue of the
exponential sensitivity to initial conditions which is the hallmark of
classical chaos.  Considerable insight into such problems has been
obtained using both semiclassical and statistical arguments.  The
semiclassical theory pioneered by Gutzwiller \cite{Gutzwiller} differs
substantially from the familiar Bohr-Sommerfeld quantization due to
the chaotic nature of the classical orbits in such systems. The main
applications of these ideas have been in atomic and solid-state
physics (see for example Ref.~14)
where detailed quantitative agreement between theory and
experiment has been achieved in a number of cases.  In each case the
theoretical work has begun from an understanding of the chaotic
classical mechanics. In perhaps the best-studied system, the hydrogen
atom subjected to intense microwave radiation, many of the puzzling
features of the variation of the ionization threshold with frequency
and intensity can be understood on a wholly classical basis, by
locating the transition point to (global) classical chaos \cite{dima}.
The theory we develop below is similar to this work in spirit,
although substantially different in detail \cite{photoatom}.  In
particular, in the present work we deal with a time-independent
energy-conserving dynamical system whose transition to chaos is
described by the Kolmogorov-Arnol'd-Moser (KAM) theory of non-linear
dynamics.  Although it is not an exaggeration to term this theory one
of the highest achievements of modern mathematical physics, it has had
very few direct applications to experimental systems; we believe that
this fact should lend interest to the experimental study of ARCs.  The
essential idea of the KAM theory and its application to ARCs will be
explained in detail in section 3 below.

\subsection{Applications of ARCs}
Before beginning the description of the theory of ARCs let us mention
the possible practical applications such resonators may have in optics
or optoelectronics.  Dielectric microspheres or microdisks already
provide compact high-Q resonators which may be useful as spectral
filters or as components in microcavity lasers.  Based on the work to
be presented below \cite{optlett}
we can propose three further advantages that ARCs
may provide:

\begin{enumerate}
\item The ability to tune
the Q-value and resonant frequency of the cavity by appropriate
deformations.

\item If this can be done {\it in situ}, then the possibility of designing a Q-switched 
ARC micro-cavity laser.

\item The ability to couple a high-Q WG mode out of the cavity with strong directionality.

\end{enumerate}
\noindent
The third item is perhaps of most importance since the lack of
intrinsic directionality in microdisk lasers has been recognized as a
major barrier to their development for applications \cite{Levi}.  
If ARCs can combine the advantages of microdisk or
microsphere cavities with the directional emission found in
Fabry-Perot resonators it will become possible to develop useful
microlasers which can take full advantage of the reduction of the
lasing threshold which arises as the number of cavity modes
overlapping the gain region is decreased \cite{Yamamoto}.  The current
generation of semiconductor microlasers (the Vertical Cavity
Surface-Emitting Laser) does not gain much of this advantage due to
the fact that the lasing modes are only two-dimensionally confined.
Of course obtaining a large threshold reduction requires a microcavity
whose dimensions are comparable to the wavelength of light and the
ray-optics model which will take up the bulk of this article does not
apply when the wavelength approaches the cavity size.  Nonetheless we
have solid numerical evidence that the qualitative behavior predicted
by the ray-optics model and in particular the directionality of
emission are still present in this limit.  For all these reasons we
believe that ARCs represent a new class of cavity resonators whose
properties and usefulness need to be explored in a series of
controlled experiments on a well-characterized realization.  As a
first step we have recently applied the theory to explain the lasing
intensity profile of deformed lasing droplets \cite{Mekis} (see
section 8), but much more work is needed and appropriate
solid ARCs need to be designed and studied.

\section{Physical Origin of Q-spoiling in ARCs}
\subsection{EM/QM Analogy}
\begin{figure}[bt]
\href{http://darkwing.uoregon.edu/~noeckel/publications/mcchapter.pdf}
{Figures available from author's website.}
\caption{
\label{fig1}
Ray trajectories for circle (a), and quadrupole-deformed circle (b) parametrized by 
$r(\phi) = 1 + \epsilon\,\cos 2\phi$ in polar coordinates for
$\epsilon = 0.08$ corresponding to an 8\% fractional deformation, or
an eccentricity $e=0.56$.
Rays are launched from the boundary at the same
$\phi$ and angle of incidence $\sin\chi_0=0.7$ in both cases; ray escape 
by refraction only occurs in case (b).
}
\end{figure}
As noted in the introduction, problems in quantum chaos are often
studied with semiclassical methods.  The analogue of the classical
limit in quantum mechanics is the ray-optics limit ($\lambda \to 0$)
of the (Maxwell) wave equation and so we begin by considering the ray
optics of a deformed cavity.  For simplicity consider an infinite
dielectric cylinder of radius $R$.  In Fig.~\ref{fig1}(a) we show the
ray motion corresponding to a typical WG mode of a {\it perfect}
cylinder in the (x-y) plane perpendicular to its axis.  Note that as
the ray circulates it always collides with the surface of the cylinder
at the same angle of incidence; this is clear geometrically, but may
also be regarded as a consequence of the conservation of the
z-component of angular momentum in a rotationally symmetric potential.
Therefore, if such a ray is trapped by total internal reflection at
its initial bounce, it will remain trapped forever within the strict
ray-optics approximation.  Of course even if we ignore for the moment
absorption or microscopic imperfections in the dielectric which may
limit the lifetime, it is well-known that such resonances do have a
finite lifetime due to evanescent leakage.  This may be seen clearly
by invoking the similarity between the wave equation in quantum
mechanics and electromagnetism \cite{Johnson}.  If we assume an index
of refraction $n=1$ outside the scatterer and $n>1$ but constant
inside the dielectric, then in each of these regions the wave equation
for the electric field has the form
\be
-\nabla^2{\bf E} +(k^2 - n^2 k^2){\bf E} = k^2{\bf E}
\ee
where $k=\omega c$ is the wavevector of the light in vacuum.  Even
though the matching conditions for ${\bf E}$ at the interface of the
dielectric can be very different from the quantum mechanical ones for
the wave function (depending on the polarization), 
one can still consider $V_{EM}({\bf r}) = (1 -
n^2({\bf r}))k^2$ as the analogue of a potential.  One sees therefore
that the dielectric body acts like an {\it attractive} potential well
for the scattering of EM radiation, and in general an attractive well
does not give rise to very sharp resonances.  However the occurence of
sharp resonances for a cylindrical dielectric can be understood by
noting that if one separates the wave equation in cylindrical
coordinates $(\rho,\phi,z)$ then the equation for the radial variation
of the electric field contains an effective potential $V_{eff}(\rho) =
V_{EM}(\rho) + m^2/\rho^2$ where $m$ is an integer representing the
conserved $L_z$ in the system.  Hence the effective potential is the
sum of the attractive well due to the dielectric and the repulsive
$L_z$ angular momentum barrier, see Fig.~\ref{fig2}. 
\begin{figure}
\href{http://darkwing.uoregon.edu/~noeckel/publications/mcchapter.pdf}
{Figures available from author's website.}
\caption{
Effective potential picture for whispering gallery resonances;
$k_{min} = m/(nR)$ and $k_{max} = m/R$. 
\label{fig2}
}
\end{figure} 
For wavevectors satisfying
\be m/(nR) < k < m/R \label{kcond}\ee
there exists a ``classically allowed'' region inside the dielectric
separated by the angular momentum barrier from the outside region
where propagation is also allowed.  Thus sharp resonances are found in
this interval of wavevector for this angular momentum channel; their
width in the ideal system is determined by the ``tunneling''
(evanescent leakage) through the angular momentum barrier.

\subsection{Classical Escape and Ray Chaos}
However this point of view makes it clear that these sharp resonances
arise {\it only} due to conservation of $L_z$.  In the same wavevector
interval there exist broad resonances corresponding to lower $L_z$; a
perturbation which destroys the rotational symmetry of the dielectric
will mix these sharp resonances with this broad continuum and
ultimately spoil their Q. The necessary mixing is not easily
calculated perturbatively because the high-Q WG modes are separated in
space from the lower Q modes and a weak perturbation will just mix
different WG modes and not spoil their Q.  Therefore it is useful to
consider the question from the point of view of ray optics.  In
Fig.~\ref{fig1}(b) we show the ray dynamics corresponding to a
cylindrical dielectric with a 8\% quadrupolar deformation in the plane
perpendicular to its axis.  The same initial angle of incidence
$\sinchi_0$ is used as in Fig.~\ref{fig1}(a) but in this case, due to
the deformation, the angle of incidence $\sinchi$ is not conserved and
fluctuates from bounce to bounce.  This raises the simple question:
will the angle of incidence fluctuate enough after a large number of
bounces so that $\sinchi$ will become smaller than $\sinchi_c=1/n$,
allowing the ray to escape refractively according to Snell's law?  If
so, then intuitively one expects the Q of the corresponding WG
resonances to be spoiled; if not there should still exist high-Q WG
resonances at this deformation.  Since a typical WG mode of the type
shown in Fig.~\ref{fig1}(a) will have a measured Q value $ > 10^8$ for
a $10 \mu {\rm m}$ diameter cross-section this implies roughly $10^6$
collisions occur (at optical wavelengths) before escape.  Therefore
this ``classical'' Q-spoiling mechanism will become relevant if escape
occurs in less than one million bounces; i.e. we are concerned here
with the very long-time asymptotics of the ray motion.  How can we
determine when such classical Q-spoiling occurs?

The ray motion in the undeformed cylinder (Fig.~\ref{fig1}(a)) is
obviously confined by a circular caustic curve (a caustic curve is a
curve which the ray touches at tangency once between each bounce).
The presence of this caustic constrains the fluctuations in $\sinchi$
(to be zero in this case).  We will see shortly that for small
deformations a dense set of caustics still exists and puts a bound on
the fluctuations in $\sinchi$ which is sufficient to prevent classical
escape.  The existence of caustic curves is characteristic of systems
which have integrable (regular) dynamics or are weakly perturbed from
integrability.  The disappearance or ``breaking'' of these caustics is
the hallmark of the hamiltonian transition to chaos for a free point
mass bouncing specularly in a hardwalled cavity (in chaos theory such
systems are referred to as ``billiards'' and when treated
quantum-mechanically as ``quantum billiards'').  The ray dynamics
problem of interest here is identical to such a billiard problem
except for the possibility of refractive escape for sufficiently small
values of $\sinchi$ (we neglect evanescent leakage henceforth unless
otherwise stated).  For our problem the breaking of the caustics
removes the bound on the fluctuations in $\sinchi$ and allows the
classical escape process and Q-spoiling of the WG modes.  Hence in
order to understand when and how this classical Q-spoiling occurs one
must look to the theory of the hamiltonian transition to chaos, which
is based on KAM theory \cite{KAM}.  The specific results on the
breaking of caustics in convex 
billiards which are relevant here are due to
Lazutkin \cite{Lazutkin}.

\section{The Hamiltonian Transition to Chaos}\label{hamchaos}
\subsection{Integrable and Chaotic Hamiltonians}

Before explaining the implications of KAM theory for our problem, we
briefly review the concepts of integrability and chaos in hamiltonian
systems \cite{Reichl}.  
A hamiltonian $H(\{ q_i,p_i\})$ is {\it integrable} if there
exists a canonical change of variables to {\it action-angle}
coordinates $\{a_i,J_i\}$ such that the new hamiltonian
$H'(\{a_i,J_i\})=H'(\{J_i\})$, i.e. the energy only depends on the
action variables $J_i$ and not on the angles $a_i$.  For generic
hamiltonians such a transformation does not exist; it is only for
systems with a high degree of symmetry that such a change of variables
can be found (unfortunately these systems represent almost all of the
examples treated in textbooks, so they appear more common than they
are).  However if such a transformation exists then it immediately
follows from hamilton's equations that all the $J_i$ are constants of
motion ($\partial H /
\partial a_i = 0$), and that the angles vary linearly with time (modulo $2 \pi$) with a 
frequency $\omega_i = \partial H /\partial J_i$.  If we specialize to
the two-dimensional case (which will apply to all the examples treated
in this chapter), the phase-space of the motion is four-dimensional,
with motion confined to three-dimensional surfaces of constant energy.
However if the motion is integrable, then in fact any given
phase-space trajectory lies on a two-dimensional subspace of the
constant-energy surface specified by the conserved values of
$J_1,J_2$.  Since the angle variables are periodic, this subspace has
the topology of a torus and the phase-space trajectory winds around
this torus with $a_1=a_1(0) + \omega_1 t, a_2=a_2(0) + \omega_2 t$.
If we define the winding number of a given torus as $W=\omega_1/
\omega_2$ then the motion differs depending on whether $W$ is a rational or irrational number.
If $W=p/q$ ($p,q$ integer) then each trajectory on that torus will
close on itself after $pq$ windings and we have an infinite family of
periodic orbits on that torus, no one of which covers densely the
entire torus.  On the other hand, if $W \neq p/q$ the orbit will never
close on itself and each orbit will cover the entire torus
quasi-periodically.  Thus for the integrable system phase-space is
divided up into nested {\it rational} and {\it irrational} tori
corresponding to these two cases.  We will see that for the relevant
integrable billiards each irrational torus defines a caustic curve.

Independent of whether the motion is on a rational or irrational
torus, if the system is integrable this motion is not ergodic, i.e. a
typical phase-space trajectory does not visit all regions of the
constant energy surface with equal density as $t \to \infty$, because
each trajectory lives on a 2D subspace of this 3D surface.  Now
suppose an additional term is added to the hamiltonian which breaks
its symmetry and destroys its integrability.  It turns out that for
sufficiently large perturbations in many cases there are no
approximate conservation laws and typical trajectories starting
anywhere in phase-space explore the entire 3D constant energy surface
ergodically.  Moreover any two trajectories corresponding to nearby
initial conditions separate from one another exponentially in time (up
to some time scale).  This situation is referred to as {\it hard
chaos}.  However KAM proved that for sufficiently weak non-integrable
perturbations there still exists some remnant of the conservation laws
present in the integrable limit and in particular that there exist
many families of trajectories confined to a 2D subspace of phase-space
(which thus are obviously not ergodic).  It is best to explain this
point in detail in the context of several examples relevant to ARCs.

\subsection{KAM billiards}\label{kamsec}

For the application to ARCs we wish to consider two-dimensional or
three-dimensional families of billiards which are generated by
deformations of circles or spheres, see 
Fig.~\ref{fig3})
\begin{figure}
\href{http://darkwing.uoregon.edu/~noeckel/publications/mcchapter.pdf}
{Figures available from author's website.}
\caption{
\label{fig3}
Integrable shapes (left) in two and three dimensions and their
non-integrable deformed counterparts.
}
\end{figure}
(the 2D case will describe
cylindrical dielectrics with deformed cross-section, the z-motion
being decoupled).  In each case the size of the deformation will be
specified by a continuous parameter which gives a circle or sphere
when set equal to zero.  We call these {\it KAM billiards} because
increasing the shape parameter from zero generates a KAM/Lazutkin
transition in the dynamics.  It is simplest to specify these billiards
by giving the radius as a function of the relevant angle ($\phi$ in
the 2D case, $\theta$ in the 3D case (where we will focus on the
systems with azimuthal symmetry).  We can think of each such
deformation as being a sum of multipole distortions, the simplest
being the dipole and quadrupole distortions, which in 2D take the
form:
\begin{eqnarray}
r(\phi) &=& \frac{1}{\sqrt{1+\epsilon^2/2}}\,(1+\epsilon\,\cos \phi)
\qquad\,\,({\rm dipole})\\
r(\phi) &=& \frac{1}{\sqrt{1+\epsilon^2/2}}\,(1+\epsilon\,\cos
2\phi)\qquad({\rm quadrupole})
\end{eqnarray}
The deformation parameter here is the constant $\epsilon$ multiplying
the dipole or quadrupole term (the prefactors insure that the area
remains constant and equal to $\pi$).  The dipole billiard was first
introduced by Robnik in order to study the quantum KAM transition
\cite{Robnik}.  Note that the dipolar distortion, to leading order in
$\epsilon$, corresponds to a translation of the center of mass of the
billiard which does not change its internal ray dynamics; when
expanded around its new center of mass the distortion is primarily
quadrupolar but without the four-fold symmetry of the quadrupole
billiard.  Since the dipole (Robnik) billiard becomes non-convex at
smaller fractional deformations it shows a higher degree of chaos (for
the same deformation) than the quadrupole billiard.  In contrast, it
is known \cite{Berry} that an exactly elliptical deformation of the
circular billiard does not generate chaotic dynamics at all: there
still exists a constant of motion, not the angular momentum with
respect to the center, but rather the product of the angular momenta
with respect to the two foci, so the motion is still integrable.  We
may use the ellipse as a ``control'' shape to test the importance of
chaos in the behavior of deformed cavities.  Finally, axially
symmetric deformations of the sphere can also be expressed in terms of
a 3D multipole expansion of Legendre polynomials $P_l(\cos \theta)$
(an example is illustrated in Fig.~\ref{fig3}); 
this type of
description will be used to model deformed liquid droplets in section
8 below.

To compare the different 2D shapes, we introduce the {\rm
eccentricity} in analogy to the ellipse,
\be
e\equiv\sqrt{r_{max}^2-r_{min}^2}/r_{max},
\ee
where $r_{max}$ and $r_{min}$ are the extremal radii measured from the
center of mass. The figures in the following sections will use this
parametrization whereas we will find it helpful in the discussion to
specify the fractional deformations in percent.

\subsection{The Poincar\'e Surface of Section}\label{poinc}
Given the shape of the billiard and the specular reflection
assumption, following the trajectory of specific rays is a simple
exercise in geometry.  However it is not always straightforward to
detect structure in the phase-space motion simply by looking at the
real-space trajectory.  On the other hand the actual phase space
trajectory typically occurs on a high-dimensional manifold which is
difficult to visualize.  Thus it is standard in non-linear dynamics to
observe the phase-space motion by imaging the particle/ray every time
it passes through a specified plane in phase-space; the resulting plot
is called a Poincar\'e {\it surface of section} (SOS).  For billiards
it is simplest to choose the SOS to correspond to the boundary, with
coordinates corresponding to the angular position $\phi$ around the
boundary (since the billiards are convex there is only one boundary
point for each angle), and the tangential component of momentum at the
bounce point which is proportional to $\sinchi$, the angle of
incidence (see Fig.~\ref{fig4}).  
\begin{figure}
\href{http://darkwing.uoregon.edu/~noeckel/publications/mcchapter.pdf}
{Figures available from author's website.}
\caption{
\label{fig4}
Definition of the coordinates for the Poincar{\'e} section of 2D billiards.
}
\end{figure}
There exists a non-linear (twist)
\cite{Reichl} map $T_{\epsilon}$ which maps the vector
$(\phi_n,\sinchi_n)$ at the $n^{\rm th}$ 
bounce to $(\phi_{n+1},\sinchi_{n+1})$
at the next bounce.  For the case of 2D billiards this map is
typically defined implicitly by the roots of a third, fourth or fifth
order polynomial.  For every trajectory the coordinates
$(\phi,\sinchi)$ are plotted at each bounce (up to some number of
bounces, e.g. 1000) giving a series of points in a 2D SOS plot (see
Fig.~\ref{fig5}); 
whether or not these points form a connected curve
depends on the nature of the phase-space motion. It is usual to plot
many trajectories in a single SOS in order to image fully any
phase-space structure which will prevent a single trajectory from
visiting all regions.  Initially we will neglect the possibility of
ray escape in discussing the SOS.

The series of SOS's shown in Fig.~\ref{fig5} for the case of the
quadrupole billiard as the deformation is increased clearly
illustrates the content and implications of the KAM/Lazutkin theory.
For zero deformation the SOS has two types of structures, solid
horizontal lines and sets of discrete points arrayed on a horizontal
line (we have denoted these sets of points by large symbols in the
figure in order to make them more visible, but each corresponds to a
single pair of bounce coordinates $(\phi,\sinchi)$). Each solid line
is the image of one quasiperiodic trajectory with irrational winding
number, since these trajectories never close after many bounces they
have visited the entire boundary (all values of $\phi$) and they yield
a full line in the SOS.  Note from the real-space picture of these
orbits that such an orbit does generate a (circular) caustic.  The
trajectories which generate discrete sets of points in the SOS are the
periodic orbits with rational winding number.  The simplest of these
are the two-bounce diametral orbits, two of which are denoted by plus
and asterisk signs in Fig.~\ref{fig5}(a).  There are an infinite family
of such orbits created by rotation of any one; each of these orbits
are marginally stable, which means that nearby (non-periodic) orbits
neither oscillate around the periodic orbit nor diverge from it
exponentially.  One such periodic orbit does not trace out a caustic
curve but simply generates an inscribed polygon of the circle (e.g. a
square for the 4-bounce orbit denoted by the squares).
\begin{figure}
\href{http://darkwing.uoregon.edu/~noeckel/publications/mcchapter.pdf}
{Figures available from author's website.}
\caption{
\label{fig5}
Poincar{\'e} surfaces of section for the circle (a) as well as quadrupolar
deformations with eccentricities $e=0.51$ (b) and $e=0.63$
(c). Vertical and horizontal two-bounce orbits in (a), shown as 
crosses and stars, are depicted in the schematic below the SOS.
Shown to the right of each SOS are trajectories starting at
$\sin\chi_0=0.7$ in all cases. In the schematics below (b), 
trajectories close to the horizontal and vertical diametric 
orbits are plotted (each below its bounce position in the SOS). 
}
\end{figure}

Fig.~\ref{fig5}(b) shows the SOS for the quadrupole billiard with $\epsilon=.065$
corresponding to a $6.5\%$ fractional deformation; in this case the
vertical diameter has been shortened and the horizontal one stretched.
Now there are {\it three} types of structures in the SOS: closed
curves which do not span the SOS (islands), grainy (chaotic) regions
which are two-dimensional and have no apparent structure, and curves
which span the SOS roughly horizontally (although they are no longer
straight horizontal lines).  The islands arise from periodic orbits
which have been stabilized by the deformation.  Stability in this
context means that nearby orbits oscillate around the central periodic
orbits.  For example, since the vertical diameter has been shortened
the vertical two-bounce orbit is now stable (this corresponds to the
well-known fact in resonator theory that two confocal spherical
mirrors separated by less than twice their radii of curvature are
stable).  Thus we see islands corresponding to nearby oscillatory
orbits around the two-bounce vertical periodic orbit [the pluses in
Fig.~\ref{fig5}(a)].

In contrast, the stretching of the horizontal diameter has
destabilized the two-bounce horizontal periodic orbit; nearby orbits
diverge from it exponentially (see inset) and give rise to a 2D
chaotic layer in the SOS.  The fact that this layer is two-dimensional
is important; no longer are single trajectories confined to points or
lines on the SOS.  This is the beginning of chaos, with exponential
divergence of nearby trajectories and single trajectories exploring a
finite fraction of the entire phase-space.  Note that both the islands
and the chaotic layers appear near {\it periodic} orbits of the
unperturbed system; trajectories in that neighborhood are not
perturbatively related to those of the undeformed billiard (a problem
which was discovered by Poincar\'e in 1899
\cite{Poincare}).  However, as
noted above, there still exist full 1D curves spanning the SOS in the
deformed case.  These correspond to the (deformed) quasi-periodic
orbits, which still trace out only a one-dimensional curve in the SOS
which is only deformed from the unperturbed straight line.  Thus these
orbits are not ergodic and do not explore a finite fraction of the
constant-energy surface.  KAM theory proves that for sufficiently weak
perturbations such quasi-periodic orbits still exist.  These unbroken
``KAM curves'' partition phase-space so that a chaotic trajectory
above a given curve in the SOS can never jump below that curve.  For
example in Fig.~\ref{fig5}(b) one sees an unbroken KAM curve with a
minimum of $\sinchi=0.5$ and a maximum value of $\sinchi
\approx 0.65$; no ray starting with $\sinchi_0 > 0.65$ can reach values of $\sinchi < 0.5$ so
for an index of refraction $n=2 \; (\sinchi_c = 1/n=0.5)$ no classical
escape is possible for a large set of WG orbits.  Thus the unbroken
KAM curves constrain the ``diffusive'' motion in the chaotic regions
to lie in layers bounded by two KAM curves.  Associated with each such
``KAM curve'' in the SOS is a caustic in the real-space motion which
prevents rays from penetrating into the interior of the cavity, as
seen in the inset.  Lazutkin \cite{Lazutkin} proved that for any 2D
convex cavity with a sufficiently smooth 
boundary such caustics exist, although as
the cavity is very deformed from the circle these caustics only
persist 
very near the boundary and may not constrain the relevant rays.

Fig.~\ref{fig5}(c) shows the SOS for slightly higher deformation,
$\epsilon=0.1$.  The increased deformation has increased the degree of
chaos in the motion and now unbroken KAM curves exist only very near
the boundary ($\sinchi > 0.9$); the bulk of the SOS is spanned by a
chaotic region.  A ray starting with $0.7<\sinchi_0 < 0.9$ will
typically be in this chaotic region and will diffuse to any value of
$\sinchi < 0.9$ over some time scale.  If as above $\sinchi_c=0.5$
then such a ray will escape according to Snell's law (an example is
shown in the inset).  A SOS for $\epsilon > 1/5$ (for which the
quadrupole is non-convex) would show no KAM curves or islands, only a
uniform sea of chaos covering the entire SOS.

\section{Ray Optics Model for Q-spoiling of ARCs}

\subsection{Classical Q factor}\label{classq}
The qualitative behavior of the ray dynamics of ARCs depicted in
Fig.~\ref{fig5} is universal for all such smooth deformations of the
circle (this is guaranteed by Lazutkin's theorem). That is why it is
possible to find general physical principles that apply to all ARCs.
What implications does this picture have for the whispering gallery
modes of ARCs?

First, intuitively one associates WG modes with rays starting at
$\sinchi_0$ close to unity (i.e. rays moving almost tangent to the
boundary).  Whereas conventional geometric optics treatments for
nonresonant scattering \cite{scoptics} start out with an incident
beam, our method deals exclusively with rays escaping from inside the
cavity. This will allow predictions about intrinsic properties of the
resonant state that do not depend on the input coupling, such as
resonance lifetime and directional lasing emission.
In the eikonal approximation for the modes of a dielectric
cylinder (see subsection 7.1) it is possible to associate an
ensemble of rays having a fixed value of $\sinchi$ with a mode of
angular momentum index $m$ and wavevector $nk$ (within the dielectric)
by the relation:

\be \sinchi = \frac{m}{nkR}. \label{semiclass1}
\ee

For the deformed cylinder no such simple relation exists (this is a
key issue in the semiclassical quantization of chaotic systems
\cite{Gutzwiller}), but as a rough starting point for a ray-optics
model of ARCs we will use Eq.~(\ref{semiclass1}) to connect the ray
dynamics to the modal properties of ARCs.  We have checked that our
main conclusions are insensitive to this assumption.  With this
assumption we can define a ray-optics procedure for estimating the
Q-factor of ARCs.

\begin{itemize}

\item {\it Assign an index $m$ to the ARC mode which evolves by continuous deformation from
the mode with angular momentum $m$ and wavevector $k$ in the
undeformed cylinder}.

\item {\it Associate this ARC mode with an ensemble of rays which is uniformly distributed in
$\phi$ but has starting angle of incidence chosen by the eikonal rule:
$\sinchi_0=m/nkR$}.

\item{\it Propagate this ray ensemble forward in time and compute the mean classical 
escape rate $\tau^{-1}$, where the classical escape probability is
zero for $\sinchi > 1/n$ and unity for $\sinchi < 1/n$}.

\item {\it Define the quality factor associated with this ray ensemble as $Q = ck \tau$}.

\end{itemize}

\subsection{Qualitative Predictions}

Our conjecture is that for dielectric ARCs with $nkR \gg 1$ the
``classical'' approximation for the Q-factor of mode $m$ is a good
means for estimating the actual width of the resonance of index $m$
for deformations that are too large to be treated perturbatively.  The
universal phase-space evolution described above allows one to make
certain qualitative predictions about the behavior of this classical
Q-factor with increasing deformation:

\begin{enumerate}

\item For modes corresponding to a given $\sinchi_0 > \sinchi_c$,
Q will be infinite (the 
escape rate will be zero) up to some threshold deformation
$\epsilon_c$ at which the last KAM curve intervening between
$\sinchi_0$ and $\sinchi_c$ breaks.  (In fact there will be some large
but finite Q due to tunneling which we neglect).

\item For deformations $\epsilon > \epsilon_c$ the Q of these modes will be finite and
decreasing as $(\epsilon - \epsilon_c)^{-\alpha}$ where the exponent
$\alpha$ is in the range $2.5 < \alpha < 3.5$.

\item Once classical escape becomes possible the emission from these modes is highly directional
and occurs (in the near field) only at the points where the regions of
chaos in the SOS connect $\sinchi_0$ to $\sinchi_c$.

\end{enumerate}

\noindent
The first prediction follows clearly from the discussion in subsection
3.3; the second and third will be justified in more detail
below.

\section{Threshold Behavior and Power Law}\label{powerlaw}

Numerical evaluation of the classical approximation for Q for any
simple model of 2D ARCs is straightforward; it just requires
simulations of the same type used to generate the SOS's in
Fig.~\ref{fig5}, except here we only start ray trajectories at the
value of $\sinchi_0$ of interest and escape is allowed at a chosen
$\sinchi_c$. The mean inverse length $L^{-1}$ of the escaping
trajectories is then computed and used to evaluate $Q= nkL$.
Representative results for the dipole and quadrupole ARCs are shown in
Fig.
\ref{fig6}. 
\begin{figure}
\href{http://darkwing.uoregon.edu/~noeckel/publications/mcchapter.pdf}
{Figures available from author's website.}
\caption{
\label{fig6}
Escape rate for the dipole and the quadrupole with refractive 
index $n=2$ and starting condition $\sin\chi_0=0.8$. The escape 
rate is proportional to $1/Q$. The plot assumes $nkR=100$; to
convert to other size parameters, multiply the vertical scale by
$100/(nkR)$. 
}
\end{figure}
As predicted, $Q^{-1}=0$ up to a threshold deformation ($3\%$ for the
dipole, $7\%$ for the quadrupole) and then rises steeply.  Note the
threshold is not universal and is lower for the dipole ARC because of
its stronger chaos compared to the quadrupole ARC for the same
eccentricity.

In Fig.~\ref{fig7} we plot the data in Fig.~\ref{fig6} on a log scale
vs. the log of the distance from threshold.  As predicted these give a
nice power law with an exponent $\alpha$ which is non-universal but in
the range $2.5-3.5$.  
\begin{figure}
\href{http://darkwing.uoregon.edu/~noeckel/publications/mcchapter.pdf}
{Figures available from author's website.}
\caption{
\label{fig7}
Power-law behavior of the escape rates in the previous figure, 
above the respective threshold deformations.
Asterisks and dashed line are the data and linear fit for the dipole,
squares and solid line refer to the quadrupole. The 
law $1/Q\propto (e-e_c)^{\alpha}$ is seen to hold, where
$\alpha_d = 2.6$ for the dipole and $\alpha_q = 3$ for the quadrupole.
}
\end{figure}
Briefly, the origin of this power law variation
of Q from threshold is the following.  In the relevant cases just
below threshold there is one last KAM curve which blocks the
phase-space flux from reaching the critical angle.  As this last
intervening KAM curve breaks with increasing deformation it leaves
behind a ``sticky'' region in phase-space known as a {\it cantorus}\cite{Reichl}
(because it has the properties of a Cantor set).  The slow traversal
of this cantorus is the rate limiting step in the classical escape
process.  Therefore the dependence of Q on the distance from threshold
is determined by the mean phase-space flux through such a cantorus.
This problem has been studied in detail and it is well-established
that this flux varies as $(\epsilon - \epsilon_c)^{\alpha}$ with
$\alpha$ in the range noted \cite{power}.  Space limitations do not
allow us to reproduce those arguments here.  However we note that the
appearance of a power-law here is not surprising as the breaking of
KAM tori constitutes a kind of critical phenomenon in phase-space,
with the scale-invariance associated with second-order transitions
\cite{Kadanoff}.

\section{Ray Directionality}\label{direc}     
As noted above, we expect the ray escape from ARCs to be directional
leading to directional emission from deformed WG modes.  Figure
\ref{fig8a} shows the directional emission from the quadrupole ARC,
obtained from a ray bundle started along $\sin\chi_0=0.83$ and
escaping at $\sin\chi_c$. The simulation was performed for two different indices of refraction
corresponding to $\sin\chi_c=0.5$ and $\sin\chi_c=0.65$.
The histograms in the figure are obtained by sorting the escaping rays
into bins according to the angle $\phi$ at which they escape; this should represent the
near-field intensity distribution.  The inset depicts the expected far-field intensity pattern
taking into account the refraction of the escaping rays.  The 
near-field histogram in both cases shows no escape whatsoever
outside two very narrow regions around $\phi=0$ and $\phi=\pi$. The
very high degree of directionality found in these simulations is
initially somewhat surprising and deserves further study.  To our
knowledge, escape directionality has not been studied in the theory of
phase space transport in Hamiltonian systems described briefly in
section 3 because the systems previously considered were closed; 
although somewhat related issues have been studied in the context of
chaotic classical scattering \cite{chaosreview}.
\begin{figure}[bt]
\href{http://darkwing.uoregon.edu/~noeckel/publications/mcchapter.pdf}
{Figures available from author's website.}
\caption{
\label{fig8a}
Directionality 
for the quadrupole at
deformation $e=0.58$. The histograms
shows the probability distribution of escape positions $\phi$,
assuming $\sinchi_c=0.5$ (solid) and $\sinchi_c=0.65$ (hollow).
The inset gives the resulting far-field directionality
for $\sinchi_c=0.5$, taking into
account refraction on exit from the dielectric. The starting condition
is $\sin\chi_0=0.83$. 
}
\end{figure}

At the most general level the occurence of preferential escape
directions and their location can be understood by simple curvature arguments.
Rays total internally reflected from a point of high
curvature are unlikely to escape at the next bounce since this will
likely occur at a point of lower curvature and $\sinchi$ will have
increased.  Since the converse is not true, rays are more likely to
escape in the regions of high curvature.  In the quadrupole these
regions are near $\phi=0,\pi$ corresponding to the emission peaks in
Fig.~\ref{fig8a}.  However we shall see that such arguments alone cannot
account for the narrowness of the peaks, their detailed structure, and their persistence in the
far-field (as shown in the inset to Fig.~\ref{fig8a}).  For example, for $\sinchi_c=0.65$
there is no ray escape precisely at $\phi=0,\pi$, the points of maximum curvature.

First let us address the far-field behavior.  Even if all rays escape very
near $\phi=0,\pi$ the emission pattern in the far-field need not be
highly directional since they might escape with very different angles
of incidence.  This does not occur since the phase space diffusion in
the $\sinchi$ direction is {\it slow}; rays approach the
critical angle in small steps and escape very near $\sinchi_c$,
leading to almost tangential emission (see Fig.~\ref{fig5}(c)).
Therefore the far-field ray pattern is just as directional as the
near-field escape histogram, but is simply rotated by $\pi/2$ (again,
see inset to Fig.~\ref{fig8a}).

\subsection{Ergodic model}

\begin{figure}[bt]
\href{http://darkwing.uoregon.edu/~noeckel/publications/mcchapter.pdf}
{Figures available from author's website.}
\caption{
\label{fig8c}
Partial surface of section, only for $\sinchi$ above the 
critical line for escape ($\sinchi_c=0.5$, shown dashed). 
The shape is the same as in
Fig.~\protect{\ref{fig8a}}. The
heavy line with arrows
is the tangent invariant curve of a corresponding ellipse
superimposed on the quadrupole SOS (see subsection \protect{6.2}). 
}
\end{figure}
As a first step to understanding the escape directionality
distribution we formulate a simple model which should correctly
represent the short-time ray dynamics near the critical angle.  In
Fig.~\ref{fig8c} we show the SOS corresponding to the escape
histogram in Fig.~\ref{fig8a} with the islands of stability shown,
but the chaotic region left empty.  One sees that there is a
relatively large chaotic region developed around $\sin\chi_c$.  Since
motion in the chaotic region is ergodic (on that region) this suggests
that as a starting point we can assume that the rays in the ensemble
starting near $\sinchi_0$ will after some time fill the chaotic region
above $\sin\chi_c$ approximately uniformly.  To test this notion we
see if the observed directionality can be reproduced by starting with
a homogeneous distribution of starting conditions above $\sin\chi_c$
and iterating the map one step (allow each ray to collide once more
with the boundary).  The probability density $P(\phi)$ for escape in
an interval ${\rm d} \phi$ will then just be proportional to the area
of the SOS mapped in one step from above to below $\sinchi_c$ and into
the interval ${\rm d} \phi$.  We call this set of assumptions the {\it
ergodic model} for directionality.  In order to recover the full
symmetry of the billiard shape, we have to repeat this with an
ensemble below $-\sin\chi_c$ (i.e. reversed starting momenta) and add
the outcomes.  The resulting approximation for the distribution
$P(\phi)$ is shown in Fig.~\ref{fig9}.
\begin{figure}[bt]
\href{http://darkwing.uoregon.edu/~noeckel/publications/mcchapter.pdf}
{Figures available from author's website.}
\caption{
\label{fig9}
Escape directionality for $\sin\chi_0=0.85$ and $e=0.58$ The broad
solid histogram is predicted by the simple ergodic model, the dotted
line results from the modified ergodic model (see text).  The heavy
line is obtained from the invariant curve model. The full ray
simulation leads to the most sharply peaked distribution (solid line).
}
\end{figure}

Although the ergodic approximation overlaps well with the peaks in the
true escape probability density, the $P(\phi)$ it predicts is clearly
much broader than the exact ray-tracing distribution. Since the
ergodic model includes the short-time effects of the varying curvature
around the boundary, this indicates that the directionality of escape
from ARCs is not simply obtained from a knowledge of curvature.
Obviously the assumption of a uniform distribution of rays filling
phase-space above $\sinchi_c$ has missed some essential part of the
physics.  What has been missed is that even in the chaotic component
there is a definite flow pattern in phase-space for short times.  This
flow pattern is close to the KAM curves of the ellipse (which is
integrable and has KAM curves at all deformations).  Examples of
such KAM curves are given in Figs.~\ref{fig8c} and \ref{fig17} 
where representative trajectories of the
quadrupole billiard are found to follow them closely.   
Since for short times motion is
along those curves most of the escape occurs near where these curves
intersect the critical line at $\sinchi_c$.  Due to this flow pattern,
many initial conditions (above $\sinchi_c$) which lead to escape in
the ergodic model can only be reached if the {\em previous} reflection
occurred with $\sin\chi<\sin\chi_c$, implying that the ray would
already have escaped before getting to the assumed starting point in
phase space. One can begin to take this into account within an
extended ergodic model by assuming zero occupation probability for initial
conditions above $\sinchi_c$ which upon one-step iteration backwards
are below $\sinchi_c$ (and uniform probability for all other initial
conditions).  Fig.~\ref{fig9} shows that this extension of the ergodic model
significantly improves the predicted $P(\phi)$ as compared to the true
distribution.
\begin{figure}[bt]
\href{http://darkwing.uoregon.edu/~noeckel/publications/mcchapter.pdf}
{Figures available from author's website.}
\caption{
\label{fig17}
Flow of three trajectories for the quadrupole at 
$e=0.57$, followed for 200 bounces each. 
The solid lines are the superimposed
invariant curves of an ellipse with the same eccentricity.
}
\end{figure}

The advantage of the ergodic model is that only one or (in the
extended version) two mapping steps need be considered. For the simple
version it is then possible to express $P(\phi)$ in terms of the
one-step map.  The exact one-step map is not usually known in simple
analytic form for KAM billiards, but we have been able to develop
reasonable approximate analytic expressions in certain cases
\cite{effmap}.  If we abbreviate $p\equiv\sin\chi$, the map can be
described by the two functions ${\bar\phi}(\phi,p)$, ${\bar
p}(\phi,p)$, giving the new position and momentum as a function of the
old variables. The map is area-preserving, i.e. the Jacobian of the
transformation $(\phi,p)\,\to\,({\bar\phi},{\bar p})$ is unity.  One
can alternatively specify the map by considering the old momentum and
the {\em new} position as given so that the dependent variables are
$\phi({\bar\phi},p)$ and ${\bar p}({\bar\phi},p)$.  The differential
probability of obtaining $\phi'$ after one mapping step applied to a
homogenous starting distribution with $p>p_c$ is
\begin{figure}[bt]
\href{http://darkwing.uoregon.edu/~noeckel/publications/mcchapter.pdf}
{Figures available from author's website.}
\caption{
\label{fig8b}
Illustration of the areas $A$ and ${\bar A}$
which end up below or come from above $\sinchi_c$, respectively, in
one mapping step.}
\end{figure}
\begin{equation}
P_1(\phi') =
\frac{1}{\vert A\vert}\int\limits_{A}\!\!d\phi\,dp\,
\delta(\phi'-{\bar\phi}(\phi,p)),
\end{equation}
where $A$ is only the region above $p_c$ that gets mapped below $p_c$ ,
$\vert A \vert$ is its area, and ${\bar\phi}$ is the new position
after one reflection.  The image of $A$ under the map, ${\bar A}$, has
the same area $\vert{\bar A}\vert=\vert A\vert$ due to area
preservation. These regions are shown in Fig.~\ref{fig8b}.
We make a transformation of variables in the integral
from $\phi$, $p$ to ${\bar\phi}$, ${\bar p}$, yielding
\begin{eqnarray}
P_1(\phi')& =&
\frac{1}{\vert{\bar A}\vert}\int\limits_{\bar A}\!\!d{\bar\phi}\,d{\bar p}\,
\delta(\phi'-{\bar\phi})\\&\equiv&
\frac{1}{\vert{\bar A}\vert}\int\limits_{{\rm min}[\,{\bar p}(\phi')]
}^{{\rm max}[\,{\bar p}(\phi')] }\!\!d{\bar p}\\&=&
\frac{1}{\vert{\bar A}\vert}\left(
{\rm max}[\,{\bar p}(\phi')]-{\rm min}[\,{\bar p}(\phi')]\right).
\end{eqnarray}
Here, ${\rm min}[\,{\bar p}(\phi')]$ denotes the smallest ${\bar p}$
within ${\bar A}$ at the final angle ${\bar\phi}=\phi'$, analogously
${\rm max}[\,{\bar p}(\phi')]$.  The boundary of ${\bar A}$ is formed
by the two curves $({\bar\phi},p_c)$ and $({\bar\phi},{\bar
p}({\bar\phi},p_c))$.  In our case, ${\bar p}({\bar\phi},p_c)$ is a
unique function of ${\bar\phi}$, and as a result ${\rm max}[{\bar
p}(\phi')]=p_c$, ${\rm min}[\,{\bar p}(\phi')]={\bar p}(\phi',p_c)$.
We therefore have the simple result that
\begin{equation}
P_1(\phi') =\cases{\frac{1}{\vert{\bar A}\vert}\left(p_c - {\bar
p}(\phi',p_c)\right)\quad {\rm for}\,p_c>{\bar p}(\phi',p_c)\cr
0\hfill {\rm otherwise}.}
\end{equation}
The distribution in Fig.~\ref{fig9} is obtained by forming the
symmetrized function $P(\phi')=[P_1(\phi')+P_1(-\phi')]/2$ .

For the extended ergodic model, we need the function $p(\phi,{\bar
p})$. Then the lobes of $(\phi,p(\phi,p_c))$ above $p_c$ delimit the
region $A$. To decide whether a given $(\phi,p)\in A$ would have come
from the region $p<p_c$ in the previous reflection, we simply invert
the momentum $p$ and ask whether the forward mapping yields a new
momentum ${\bar p}$ smaller {\em in magnitude} than $p_c$. This is
true for a region $B$ bounded above $p_c$ by the curve
$(\phi,-p(\phi,-p_c))$. The modified starting domain for the ergodic
model is then $A' = A\setminus B$. 

Whereas reasonable agreement with the near-field directionality of the
full simulation can be obtained in the extended ergodic model, we do
not expect correct results for the far-field which is seen in
Fig.~\ref{fig8a} to display the same degree of directionality. Since
the ergodic models yield a final distribution of escaping trajectories
that scatters widely in $\sinchi$ (see ${\bar A}$ in Fig.~\ref{fig8b}), 
any near-field directionality will be spread out after 
refraction is taken into account. We thus need to understand why in
fact only $\sinchi\approx\sinchi_c$ occurs upon escape. 

\subsection{Tangent Invariant Curve Model}\label{tangic}
The ergodic models discussed above have limited
validity because they do not take fully into account the detailed
structure of phase-space motion.  That is the existence of a dominant
flow pattern over many bounces and the persistence of stable islands
and unstable periodic orbits (which repel phase-space flux).  If there
is a stable island which is intersected by the critical line
$\sinchi_c$ then no classical escape can occur in the interval of
$\phi$ within the island.  For example, for the particular shape
giving the SOS of Fig.~\ref{fig8c} there exists two large islands centered at
$\sinchi \approx 0.65, \phi = 0,\pi$.  These islands completely block escape at
the points of highest curvature $\phi=0,\pi$, giving the surprising peak-splitting 
shown for this case in Fig.~\ref{fig8a}.  
The ergodic model on the other hand
would find escape at those points from starting rays which could never
have gotten into the island in the first place.  The blocking of ray
escape by remaining stable islands is a major effect in the lasing
droplets which we study in section 8.  There are also interesting
effects in the directionality associated with unstable fixed points,
but we cannot discuss them in detail here.  These effects arise from
stable and unstable periodic orbits and are less universal since
different ARCs will have substantially different important orbits;
however they can be understood on an individual basis if one is
interested in a particular ARC.

Nonetheless there is still one
further universal feature which relates to the directionality
distribution of all ARCs which have a large quadrupolar component to
their deformation.  As noted above, the ellipse, which is integrable,
is for small deformations primarily a quadrupolar deformation of the circle.  Therefore all
ARCs with a large quadrupolar component generate short-time ray
dynamics close to the invariant curves of the ellipse as shown in
Fig.~\ref{fig17}. 
The existence of chaos in these ARCs allows motion
perpendicular to these curves which would be forbidden in the ellipse,
but for reasonable deformations this motion is rather slow.  Thus,
even when the SOS looks rather structureless and chaotic in the
vicinity of $\sinchi_c$ one finds as shown in Fig.~\ref{fig17} 
that the ray dynamics over hundreds of
bounces follows the nearest invariant curve (IC) of the ellipse.  Thus
a reasonable representation of the motion is slow diffusion between
these ICs until the IC which is tangent to the critical line is
reached and then rapid escape near the points of tangency of this
IC (which we call the tangent IC).  This explains why most escaping rays
do so at $\sinchi \approx \sinchi_c$.  This description will only be good
if $\sinchi_0$ is separated from $\sinchi_c$ by more than the width of
the IC in $\sinchi$, but this is typically the case for the WG
modes of interest to us here.  This picture of the phase-space motion
can be tested by comparing $P(\phi)$ for our standard starting
conditions to the $P(\phi)$ obtained by starting a uniform
distribution of points on the tangent IC.  In the latter case of
course almost all the rays escape after a few tens of reflections,
whereas many thousands are required in our ray-optics model.  One
finds that the tangent IC starting conditions reproduce the ray-optics
$P(\phi)$ with remarkable accuracy (see Fig.~\ref{fig9}).  This finding has
strong implications.  First it implies that the directionality
distribution of emission from the WG modes of ARCs is a short-time
property, insensitive to the many thousands of reflections needed to
reach the vicinity of the tangent IC.  Therefore the directionality
distribution is approximately independent of the initial conditions
over a wide range of WG trajectories.  {\it This strongly suggests
that all deformed WG modes will tend towards the universal
directionality distribution determined by the tangent IC at a given
index of refraction}.  If this is the case then the directions of
highest emission from ARCs will be independent of the particular
resonant mode excited, a property which should be quite important for
applications, e.g. in a multi-mode laser.

\section{Wave Properties of ARCs}
\subsection{Eikonal Theory for Undeformed Case}\label{semicl}

The ray optics model which we have developed in the previous sections relies on an
eikonal (semiclassical) prescription introduced in subsection 4.1
 to relate the
mode indices of a WG resonance to the initial $\sin\chi$ of a ray
ensemble.  Unfortunately, because the semiclasical quantization of a mixed phase-space is
still an open problem \cite{bohigas}, 
we only have such a rule for the undeformed case; we now explain the 
origin of this rule.  Since normally the WG resonances are very narrow even in the deformed case
a good starting point should be the spectrum of the isolated cavity (i.e. with perfectly
reflecting walls).  The actual resonances in the leaky dielectric are
frequency-shifted with respect to the isolated modes but the resulting corrections to $\sin\chi_0$ 
become negligible for large $kR$.  The WG modes have been quantized semiclassically
\cite{keller} for circular or elliptical cylinders, and for spheres.
The method employed in that work can be also be applied to
deformed cavities as long as the ray dynamics still exhibits a dense set of
caustics, as shown in Fig.~\ref{fig10}(a), however we do not attempt such an extension here.
The eikonal approximation attaches to each ray
a set of wavefronts; to describe the phase of these wavefronts we need to find the 
{\it eikonal} ${\sf S}({\bf r})$
such that $\nabla{\sf S}$ gives the ray direction at all points ${\bf
r}$. Assume a dense set of caustics of the ray motion exist, that is closed curves which the
ray touches at tangency once between each reflection from the boundary.
\begin{figure}[bt]
\href{http://darkwing.uoregon.edu/~noeckel/publications/mcchapter.pdf}
{Figures available from author's website.}
\caption{
\label{fig10}
(a) Integration path leading to Eq.~(14), shown as the jagged solid
line around the caustic C (dotted). Wavefronts of ${\sf S}_{CB}$ 
(dotted arcs) are
perpendicular to all the rays (dashed straight lines) between C and
the boundary B. The continuation of the rays through C would not be
normal to the wavefronts shown, but instead to their mirror images. 
(b) Exact resonance
positions (dots) for the circular cylinder. The heavy line belongs to
$m=kR$. The inset to (b) shows the scattered intensity from a plane
wave into the angle element at $50^{\circ}$ from the beam axis, in the
size parameter interval delimited in the main figure by horizontal
dashed lines. The index of refraction is $n=2$.  }
\end{figure}
There are two rays going through every ${\bf r}$ in the region
between caustic (C) and boundary (B), namely one coming from C to B
and one reflected from B toward C. The eikonal expression for the electric field is 
a sum of two wavefronts,
\begin{equation}\label{eikonal}
E({\bf r}) = A_{BC}({\bf r})\,e^{ink{\sf S}_{BC}({\bf r})} + A_{CB}({\bf
r})\,e^{ink{\sf S}_{CB}({\bf r})}.
\end{equation}
When following a ray between two reflections from B, we move first
perpendicular to the surfaces of ${\sf S}_{BC}=c$ ($c={\rm const}$),
until we reach the caustic. Then we get from C back to B with the wave
fronts ${\sf S}_{CB}=c$. In Fig.~\ref{fig10}(a) only one half of this
ray path is depicted for a number of starting points. 
In both halves of the ray path, the value of the phase $c$
increases by the length $l$ traveled along the ray between B and C,
because for any two points ${\bf r}_1$ and ${\bf r}_2$
\be\label{path}
c({\bf r}_2)=c({\bf r}_1) + \int\limits_{{\bf r}_1}^{{\bf r}_2}
\nabla{\sf S}({\bf r})d{\bf r}.
\end{equation}
Using the latter relation, one obtains the desired quantization
condition by following an alternative integration path that always
runs perpendicular to the wave fronts of, say, ${\sf S}_{CB}$ and in
addition forms a closed loop: Starting at ${\bf r}_1$ we can move {\em
on} the wavefront to the caustic without accumulating any phase in
(\ref{path}). We then follow a ray for an infinitesimally small length
to a neighboring wavefront of ${\sf S}_{CB}$ and move back to the
caustic along this new wavefront, see Fig.~\ref{fig10}(a).  
Repeating this procedure, one
eventually encircles the caustic without ever crossing it, accruing a
total phase equal to $L_C$, the length of the caustic, before reaching
the end point ${\bf r}_2={\bf r}_1$. In order for the function $E({\bf
r})$ in Eq.~(\ref{eikonal}) to be single valued at ${\bf r}_1$, this
phase difference must give unity when put in the exponential, i.e. we
require
\begin{equation}\label{quantc}
nkL_C = 2\pi m\quad (m=0,\pm1,\pm2,\ldots).
\end{equation}
For a circular domain of radius $R$, the length of the caustic is
related to the classical angle of incidence $\chi$ by $L_C=2\pi
R\sin\chi$. Combining these results, we arrive at
\begin{equation}\label{semiclquan}
\sin\chi=\frac{m}{nkR},
\end{equation}
which is the semiclassical expression we have used above. It can be seen that the
integer $m$ is precisely the angular momentum quantum number of
quantum mechanics if we recall that the angular momentum of the photon
at each bounce from the boundary is $M={\bf r}\times (\hbar n{\bf k})
= R\hbar n k \sin\chi = m \hbar$. Note also that a large number of
resonances -- those with approximately the same ratio $m/kR=n\sin\chi$ --
are described by the same classical ray dynamics.  Equation
(\ref{semiclquan}) can now be inserted into the condition for total
internal reflection, $1>\sin\chi>1/n$, to obtain
\begin{equation}\label{semicl1}
n>\frac{m}{kR}>1.
\end{equation}
This is identical to the condition given in Eq.~\ref{kcond}, and agrees well 
with numerical solutions of the wave equation for the undeformed case.
This is illustrated in Fig.~\ref{fig10}(b) and (c); 
the only sharp resonances observed in the open
cavity are those satisfying $kR <m$. Figure \ref{fig10}(c) shows the
resonances in a window of $kR=10\ldots 12$. Comparing with 
Fig.~\ref{fig10}(b) where the same window is indicated
by the two horizontal dashed lines, we see that the three narrowest
resonances correspond to the three points with largest $m/(kR)$; and
in general, the larger $m/kR$, the narrower the resonance. 

The standard eikonal approximation does not allow us to calculate the width in the open cavity
which is a tunneling effect.  However it is possible to develop a WKB-type approximation
for the small tunneling width in the limit of
large $kR$ but at constant $m/kR$  \cite{noeckel1}:
\begin{equation}\label{width4}
\delta k=\frac{1}{2}\,
\frac{1}{n^2-1}\,\exp[-(2m-1)(\alpha-{\rm tanh}\alpha)],
\end{equation}
where ${\rm tanh}\alpha\equiv\sqrt{1-(kR)^2/{m^2}}$ is a constant
between zero and one if we consider only rays of a fixed $\sin\chi$ (similar expressions are
known for the Mie resonances of spherical dielectrics \cite{Johnson}).  As we go to higher $kR$ 
and $m$, the width is seen to decrease exponentially. In order to understand the Q of ARC 
resonances we now ask how this narrow width changes as the cylinder is deformed.

\subsection{Exact solutions for the deformed cylinder}\label{exact}
\subsubsection{Resonance widths}In the inset to Fig.~\ref{fig10} it
was assumed that a plane wave is incident normal to the cylinder. To
solve this scattering problem, one uses polar coordinates $r,\phi$ and
performs a partial wave analysis, decomposing the incoming plane wave
into
\begin{equation}
e^{ikx} = \sum\limits_{m=-\infty}^{\infty}(-i)^m e^{im\phi}J_m(kr).
\end{equation}
Then one can solve the resulting decoupled radial equations for each
$m$. These equations are obtained by imposing the matching conditions
for the fields at the dielectric interface, noting that the field
inside must be a superposition of Bessel functions $J_m(nkr)$, whereas
the outgoing scattered wave consists of Hankel functions
$H_m^{(1)}(kr)$. One can now try to satisfy the same matching
conditions for the inside and outgoing waves alone, {\em leaving out}
\begin{figure}[bt]
\href{http://darkwing.uoregon.edu/~noeckel/publications/mcchapter.pdf}
{Figures available from author's website.}
\caption{
\label{fig11}
Exact results for resonance positions and widths in the dipole
as a function of
deformation $e$. Left column: azimuthal mode index $m=20$, radial node
number $n_r=0$; right column: $m=34$, $n_r=1$ (indices refer to
$e=0$).  Top : resonance positions; bottom: resonance widths
(logarithmic scale). The solid lines are the widths as calculated from
ray-optics simulation. }
\end{figure}
the incident wave. This is possible only at discrete {\em complex}
values of the wavenumber, $k-i\kappa$ [$\kappa>0$ if we assume a time
dependence $e^{-i\omega t}$ in the wave equation], and the
corresponding solutions are the quasibound states of the cavity. For symmetric cavities 
and perturbative deformations their
role in microcavity optics has been studied in detail previously (see
for example Ref.~9. Resonances of 
width $\kappa$ occur at wavenumber $k$ when a quasibound state exists
close to the real axis at $k-i\kappa$.

When the rotational symmetry is broken, the matching equations for
different $m$ become coupled so that we have to solve a progressively
larger system of simultaneous equations for the coefficients of the
$J_m(nkr)$ inside and $H_m^{(1)}(kr)$ outside of the cavity. However the
resulting quasibound state solutions evolve continuously out of those
for the unperturbed cylinder as the deformation increases. This
simplifies the search for resonances in the complex plane because we
can start at the known quasibound state positions of the circular
cylinder and then follow each solution as it moves through the complex
plane when we change the shape.  Representative results for the real
positions and widths of WG resonances ar shown in Fig.~\ref{fig11}.
The size parameters $kR$ we have studied at this point are not large enough to expect the
resonances to be accurately described by the ray-optics model; nonetheless the figure
shows clearly a threshold behavior in the resonance width as expected
from the classical model.  The width remains small and almost constant
up to a critical deformation, beyond which a rapid broadening is found. 
One also sees that above the classical threshold the ray-optics
model appears to give order-of-magnitude agreement with the width (note that there are no
free parameters in this comparison).  However for the resonances
treated in Fig.~\ref{fig11} this broadening begins before the classical Q-spoiling threshold
\begin{figure}[bt]
\href{http://darkwing.uoregon.edu/~noeckel/publications/mcchapter.pdf}
{Figures available from author's website.}
\caption{
\label{fig12}
Comparison of exact wave results for dipole (crosses), quadrupole
(asterisks) and ellipse (circles). The resonance considered is $m=20$,
$n_r=0$ (in the undeformed case). The arrows indicate the classical
thresholds for dipole (D) and quadrupole (Q).  }
\end{figure}
deformation is reached and has an approximately exponential dependence on the distance from
threshold (and not the power-law predicted by the ray-optics model).  We attribute this
behavior to tunneling effects which will always be important near threshold because rays
approach closely the critical angle.  This would explain both the reduction of the threshold
deformation and the exponential-like behavior (note that
Eq.~(\ref{width4}) 
predicts exponential variation
of the width with $\sinchi$ in the undeformed case).

Clearly at these values of $kR$ tunneling and interference corrections to the ray-optics model
must be included.  Such corrections then lead us into the true subject matter
of quantum or wave chaos, and much more work needs to be done to reach a full 
semiclassical understanding of mixed phase-space quantum systems.  Nonetheless we
expect that as $kR$ increases these non-classical effects will become weaker and the
ray-optics predictions will be more closely obeyed.   For example, in the droplets
studied in section 8, 
$kR$ is at least an order of magnitude larger than in Fig.~\ref{fig11}.
The results of Fig.~\ref{fig11} do however raise the question of whether at the moderate
size parameters $kR \sim 10-25$ investigated the classical mechanics of the cavity is relevant
at all to the $Q$ of its WG resonances.  In order to answer this question we compare in
Fig.~\ref{fig12} the resonance widths of the three different types of cylindrical deformations 
introduced so far: elliptical (which is integrable), quadrupolar which exhibits a KAM transition to
chaos, and the dipolar (Robnik) deformation which at equal fractional
deformation exhibits more chaos than the quadrupole, as discussed in
section 3.  In all three cases a sort of threshold behavior is seen with initial
exponential broadening above threshold.  However the slope of this broadening differs substantially
in the three cases with the most chaotic dipole ARC showing the most rapid broadening, the
less chaotic quadrupole ARC having an intermediate slope, and the non-chaotic elliptical ARC
showing the slowest broadening.  At an eccentricity of $0.5$ the dipole and elliptical WG modes
corresponding to the same undeformed resonance differ in width by almost two orders of magnitude!  
This suggests that the rapid breakup of caustics which occurs in the dipole ARC and does not occur
at all in the elliptical ARC has an important influence on the Q value of their resonances even
at these moderate wavevectors.

\subsubsection{Wave directionality}
To test the predictions of section 6 for the directionality
of emission from a deformed cavity, we have to choose the proper
boundary conditions in the wave equation corresponding to the situation
of a uniform ray ensemble starting {\it inside} the dielectric and then escaping. This
emission process differs from a scattering experiment which requires
an incoming wave to excite the resonance. The directionality pattern
in a scattering experiment will depend on the form of the incident
wave both because of interference with the outgoing wave, and because the incident wave
may couple preferentially to different senses of circulation of the rays.
These effects are absent in emission, so a unique directionality profile will be
observed that depends only on the quasibound state itself and should be
approximately described by our ray optics model if the size parameter is sufficiently large.

If the resonant state is at the complex frequency
$\omega-i\gamma\equiv c(k - i\kappa)$, then the corresponding solution
of the time dependent wave equation decays at a rate $\gamma$ since it
has the form $E({\bf r},t)=E({\bf r})\,e^{-i\omega t}\,e^{-\gamma t}$
where $\gamma>0$. But as a function of $r$, the outgoing waves in fact
exhibit exponential growth because 
\begin{equation}\label{asymp}
H_m^{(1)}(x)\approx\sqrt{\frac{2}{\pi x}}\,e^{i(x-m\pi/2- \pi/4)}
\end{equation}
for large values of $x =(k-i\kappa)r$.  The physical reason for this
growth with $e^{\kappa r}$ is a retardation effect: the field at $r\gg
R$ has propagated away from the cavity where it originated a time
$\Delta t\approx r/c$ in the past -- but at that earlier time the field at the
cavity was larger by a factor $e^{-\gamma\Delta t}$. This is equivalent to $e^{\kappa r}$.

As can be seen from Eq.~(\ref{asymp}), all the Hankel functions in the
outgoing wave depend on $r$ through the {\em same} factor
$\sqrt{\frac{2}{\pi x}}\,e^{i(k-i\kappa)r}$ in the far-field ($r\gg
R$).  Pulling out this common dependence, the field of the quasibound
state factorizes into radial and angular functions,
\begin{equation}\label{factor}
E({\bf r}) = \sqrt{\frac{2}{\pi x}}\,e^{i(k-i\kappa)r}\,E(\phi).
\end{equation}
This means that the directionality at large distances becomes
independent of $r$, being contained solely in $E(\phi)$. We choose $r$
in this far-field region and plot the square of the electric field
(which is proportional to the intensity) as a function of $\phi$ to
obtain the wave directionality. Figure \ref{fig13} shows the intensity
profile for a resonance of the quadrupole and compares it to the classical directionality
histogram for the corresponding ray ensemble.  One sees quite reasonable agreement between
the full widths of the intensity maxima in the two approaches.  The wave intensity distribution
is modulated by interference effects which are neglected in the pure ray-optics model (but
which should be present in a semiclassical treatment).  Moreover the wave-intensity envelope
does not show the peak splitting found in the ray-optics calculation.
It will be interesting to see if resonances at larger $kR$ corresponding to
the same $\sinchi$ begin to show this peak splitting in their envelope.
\begin{figure}[bt]
\href{http://darkwing.uoregon.edu/~noeckel/publications/mcchapter.pdf}
{Figures available from author's website.}
\caption{
\label{fig13}
Far-field directionality of light emitted from the quadrupole-deformed cylinder,
assuming a refractive index $n=2$ and $\sin\chi_0=0.83$.  The
far-field intensity is plotted as a function of direction at $e
=0.58$. The upper curve represents the ray-optics result; the lower
curve (offset for clarity) is obtained from exact numerical solution
of the wave equation (same resonance as in previous figure).}
\end{figure}

\section{Thresholds and Intensity Distribution in Lasing Droplets}
\label{droplets}
In this section we apply the ray-optics model for ARCs to the description of the WG modes
of deformed liquid droplets and give an explanation for the observed anisotropy of lasing
emission from droplets, shown in Fig.~\ref{fig14} \cite{Mekis}. In this
experiment a stream of ethanol droplets containing Rhodamine-B dye
with average radius $\approx 30\mu{\rm m}$ is
created at the vibrating orifice of a Berglund-Liu generator \cite{qian}. At the orifice the
droplets are highly non-spherical and as they fall they undergo 
damped oscillations between oblate and prolate configurations (Fig. ~\ref{fig14}) driven by excess 
surface tension until they relax to highly spherical shape far downstream. The period of the shape 
oscillations is of order
$50\,\mu s$, which is far longer than the lifetime of the WG resonances of $\approx 10\,ns$
\cite{zhang}. Thus we may treat the different phases of the droplet
oscillation as static examples of oblate, spherical or prolate microcavity lasers
and analyze the angular emission intensity in terms of the theory of ARCs.
Previously we have focused on cylindrical ARCs deformed perpendicular to their axes; here we apply
the ray-optics model to dielectric spheres deformed so as to preserve azimuthal symmetry (the
deformed droplets retain this symmetry as well as reflection symmetry through the equator to
a good approximation).

As seen in Fig.~\ref{fig14}, laser emission is fairly isotropic in the sphere, but gets
suppressed near the poles of both the oblate and prolate droplets.
Furthermore, the oblate shape is brightest around the equator whereas
the prolate shape of the same deformation (ratio of long to short
axes) emits most strongly from regions around $\theta\approx
30^{\circ}-45^{\circ}$. These observations are stable over a range of prolate and oblate
shapes (exceeding a certain degree of deformation) and are independent of the direction
of the optical pumping.  It is important to note that the three
total-energy images are normalized individually with white referring
to the maximum emission, which is of different magnitude in each
shape.  
\begin{figure}[bt]
\href{http://darkwing.uoregon.edu/~noeckel/publications/mcchapter.pdf}
{Figures available from author's website.}
\caption{\label{fig14}
Shadow graphs (a) and simultaneous total-energy images (b) of three
lasing droplets falling in air taken at different phases of
oscillation: prolate (top), spherical (middle) and oblate (bottom).
Light regions in (b) indicate lasing.}
\end{figure}

Since axial symmetry is preserved for the droplets their
instantaneous shape can be specified in spherical coordinates by
$r_b(\theta)$, independent of azimuthal angle $\phi$. The damping is
weakest for the low multipole components of the oscillation, so we
include only the Legendre polynomials $P_0(\cos\theta)$,
$P_2(\cos\theta)$, $P_4(\cos\theta)$ in an expansion of the shape. Odd
orders do not appear because of the approximate symmetry
$r_b(\theta)=r_b(-\theta)$ noted above. 
The particular shapes we use to model this
behavior more realistically are
\begin{equation}\label{shape3d}
r(\theta) = 1 + \epsilon\left(\cos^2\theta +
\frac{3}{2}\cos^4\theta\right)
\end{equation}
for prolate deformations; an oblate shape with the same axis ratio
[equal to $(5/2)\epsilon$] is obtained by replacing $\cos\theta$ with
$\sin\theta$.
 
The ray dynamics analysis is facilitated by the axial symmetry of the
droplets which implies (in the language of particle trajectories) that
the $z$ component of angular momentum, $L_z$, is conserved. At any
given $L_z$ and total energy $E$, the equations of motion thus have
only two degrees of freedom, just as in the deformed cylinder.
This becomes explicit in cylindrical coordinates $\rho$, $\phi$, $z$ where one has
\begin{equation}\label{energy}
E=\frac{1}{2}m\,({\dot\rho}^2+{\dot z}^2)+\frac{L_z^2}{2m\rho^2}.
\end{equation}

Let us look at the dynamics projected into the 2D $(\rho,z)$ coordinate system.
Each specular reflection causes a discontinuous change in ${\dot\rho}$ and ${\dot z}$; however
the angular velocity ${\dot\phi}$ remains unchanged
because the normal to the surface of an axisymmetric cavity is always perpendicular to
the $\phi$ direction. Thus a 3D specular reflection
simply reverses the normal component of the 2D projected velocity
$({\dot\rho},{\dot z})$ and reflections are also specular in the projected coordinates.  
Reflections occur whenever the trajectory
$\rho(z)$ intersects the boundary curve $\rho_b(z)$.  Between reflections the particle motion 
is free, ${\dot z}={\rm const}$, and Eq.~(\ref{energy}) can be
integrated to find $\rho(t)$. It can be shown that $\rho^2(z)$
describes a parabola whose vertex is the point of closest approach to
the $z$-axis and whose intersections with the squared boundary curve
$\rho^2_b(z)$ are the collision points. The curved trajectories in the
$z$-$\rho$-plane between specular bounces [see the inset to
Fig.~\ref{fig15}(a)] are to be contrasted with
the straight paths in conventional 2D billiards where the centrifugal
potential $L_z^2/(2m\rho^2)$ is absent. For this reason we call this
new class of systems {\em centrifugal billiards}.

\begin{figure}[bt]
\href{http://darkwing.uoregon.edu/~noeckel/publications/mcchapter.pdf}
{Figures available from author's website.}
\caption{\label{fig15}
Poincare surfaces of section for prolate droplets with deformation of
$\epsilon = 0.2$ for $L_z=0.735$ (a), $L_z=0.6$ (b), $L_z=0.45$ (c) 
and $L_z=0$ (d). The 
dash-dotted line denotes the critical line for escape, 
$\sin \chi_c =1/n =.735$ corresponding to the experimental value of
$n=1.36$ for the refractive index of the droplets in 
Fig.~\protect{\ref{fig14}}.  
Inset in (a) shows the droplet shape in the $z$ - $\rho$ plane 
and special (periodic) trajectories for $L_z=0.735$ (solid line)
and $L_z=0.2$ (dashed line).
}
\end{figure}
To discuss the resulting dynamics we introduce dimensionless variables
in Eq.~(\ref{energy}) by setting $E=1/2$ and $m=1$. Then one has
\begin{equation}\label{energy1}
1={\dot\rho}^2+{\dot z}^2+\frac{L_z^2}{\rho^2}
\end{equation}
where $0\le L_z\le\rho_b(z_{max})$ if the maximum distance from the
$z$-axis is $\rho_b(z_{max})$. To simplify notation we assume that the
droplets have their widest transverse cross-section in
the equatorial plane, i.e. $z_{max}=0$.  Again the escape condition
is simply $\sin\chi<\sin\chi_c$, where $\sin\chi$ is the angle of incidence with
respect to the surface normal ${\bf n}$ at the reflection point.  This
is not the same as the normal angle in the $\rho-z$-plane, as can be
seen by considering a trajectory reflecting entirely in the equatorial
plane at nonzero $\sin\chi$; its apparent angle of incidence in the
$\rho-z$-plane will be zero. In our units $\cos\chi = {\bf n}\cdot{\bf v}/\vert
v\vert=n_{\rho}{\dot\rho}+n_z{\dot z}$ since the total velocity is
$v=\sqrt{2E/m}=1$. The angle in the $\rho-z$-plane is then given by
$\cos\chi_{\rho z}=\cos\chi/\sqrt{{\dot\rho}^2+{\dot z}^2}$. 
It is convenient in the plotting of Poincar{\'e} sections to use as variables the polar angle
$\theta$ and the 3D $\sin\chi$ at each reflection since in these coordinates
the escape condition is still satisfied along a horizontal straight line.

At nonzero $L_z$ certain regions of the SOS are forbidden due to the
$L_z$ angular momentum barrier (e.g. a ray reaching the pole
($\theta=0$) must have $L_z=0$). 
For the allowed bounce coordinates $\theta$, $\sinchi$ one finds  
the inequality 
$\sin\chi\ge L_z/\rho_b(z(\theta))$, 
where $z(\theta)=r_b(\theta)\sin\theta$; this relation delimits the empty spaces in 
the SOS's of Fig.~\ref{fig15}(a-d) which were made for a prolate shape of fixed deformation
($\epsilon = 0.2$) and varying values of $L_z$.  Before discussing ray escape in the deformed
droplets it is important to note that as we proceed from higher to lower $L_z$ in
Fig.~\ref{fig15}(a-d) in addition to the excluded regions of the SOS decreasing (because the
angular momentum barrier becomes weaker) the degree of chaos grows rapidly.  There is actually
no visible chaos in Fig.~\ref{fig15}(a) and a mostly chaotic SOS for $L_z=0$ (Fig.~\ref{fig15}(d))
{\it for a droplet of fixed deformation}.  The reason for this is that high $L_z$ trajectories
are confined near the equator and a cross-section of the droplet at the equator is perfectly
circular, i.e. high $L_z$ orbits see an effective deformation which is much weaker than
polar orbits ($L_z=0$) which travel in the most deformed cross-section of the droplet.  
The effective deformation varies approximately as $\epsilon_{eff}= 
\epsilon\sqrt{1-L_z^2/\rho_b^2(0)}$
and tends to zero at the maximum allowed value of $L_z$.  Thus as long as $\epsilon$ is large
enough to induce classical Q-spoiling for the $L_z=0$ orbits of interest, by looking at different 
$L_z$ values for a fixed deformation one can study the classical Q-spoiling transition in a single 
ARC.  We have illustrated this situation in Fig.~\ref{fig15}(a-d).

Note that there is an absolute minimum 
allowed $\sin\chi\equiv\sinchi_m$ which occurs at the equator
($\theta=\pi/2$) where $\rho_b$ is maximal (i.e., $\sinchi_m=L_z/\rho_b(0)$). 
This implies that classical ray escape is entirely forbidden due to the angular momentum barrier
for values of $L_z\ge\rho_b(0)\sin\chi_c$; such a case is shown in Fig.~\ref{fig15}(a).  
As just noted these high $L_z$ modes are confined to orbits near the plane of the equator
[see also the inset to Fig.~\ref{fig15}(a)];
since classical escape is forbidden for these modes we always expect to find high-Q WG modes in
the equatorial region of axially-symmetric deformed microspheres.  Since this follows simply
from $L_z$ conservation it will be true in both the oblate and prolate shapes.

Proceeding now to lower $L_z$ in Fig.~\ref{fig15}(b) we see that the angular momentum barrier 
has weakened enough that the allowed region of the SOS passes through $\sinchi_c$ and rays with
this value of $L_z$ can escape.  However as before WG modes will be associated with rays starting
at large $\sinchi \approx 0.9$ in this case.  These rays are unable to reach $\sinchi_c$ due
to remaining KAM curves just as we saw earlier in our discussion of dielectric cylinders.
Therefore we expect high Q WG modes for this value of $L_z$ as well.  This situation persists
all the way to $L_z=0$ for deformations less than roughly 5\% of the radius, so we expect little
Q-spoiling and approximately isotropic emission for smaller deformations than this.

However for the 50\% deformation used in Fig.~\ref{fig15}(a-d) reducing $L_z$ a little more
causes the appearance of regions of chaos which extend from high $\sinchi$ across $\sinchi_c$
allowing classical Q-spoiling of the WG modes.  We expect all modes with $L_z$ less than this
value to have their Q rapidly degraded.  As the Q of these modes decreases it will fall
below the threshold Q-value to support lasing and these modes will go dark.  But these low $L_z$
modes are the only ones which can emit from the polar regions because of the angular momentum
barrier for the high $L_z$ modes.   Therefore our model explains naturally why the polar regions
are dark while the droplet still lases.  The low $L_z$ modes which could emit from the poles have
too low Q to lase and the high Q modes which support lasing are confined away from the polar regions.
This argument holds for both the oblate and prolate deformations in agreement with observations.

We are left with the question of why the emission profiles are
nonetheless so {\em different} in prolate versus oblate shapes.
To answer this question we must look at where the stable islands which block chaotic escape
occur for the two types of deformations.  The prolate shape corresponds to a stretching of the
droplet in the vertical direction and a compression in the equatorial plane.  Because
it is compressed in the equatorial plane there exists a large stable island 
at $\theta=\pi/2$ corresponding to
the two-bounce diametral orbit of the type we discussed in the 2D case in subsection 3.3.
This island appears (distorted due to the $L_z$ barrier) clearly in Fig.~\ref{fig15}(c,d).
As discussed above in 
subsection 6.2, a stable island intersecting the critical
line will prevent the classical escape in the corresponding 
directions; thus in the case of Fig.~\ref{fig15}(c) escape is blocked over
an interval in polar angle centered at the equator.

In the oblate droplet the situation is reversed.  The polar diameter is compressed and the
equatorial diameter is stretched and the stable two-bounce orbit (if it is still stable)
would appear at $\theta=0,\pi$.  But for most $L_z$ the island around this orbit is unreachable
and it has little effect on the dynamics.  In Fig.~\ref{fig16}(a) and (b) the SOS 
for the prolatee and oblate shapes are compared at equal deformation and equal ratio
$L_z/\rho_b(0)$.  Indeed in the SOS for the oblate shape the regular island
centered on $\theta=\pi/2$ and $\sinchi_m$
is absent because equatorial orbits with low $\sinchi$ are now unstable.
\begin{figure}[bt]
\href{http://darkwing.uoregon.edu/~noeckel/publications/mcchapter.pdf}
{Figures available from author's website.}
\caption{\label{fig16}
Prolate (a) and oblate (b) droplet SOS's at $\epsilon=0.2$ and
$L_z/\rho_b(0)=0.3$. As in the previous Figure, the dash-dotted lines
indicate $\sinchi_c$.
The escape directionality is shown in (c) for
the prolate (filled histogram) and oblate (white) shape.
}
\end{figure}
One sees no effect of the stable islands
at $\theta=0,\pi$ due to the angular momentum barrier.  There is still
an island near the critical line at $\theta=0$ and $\sinchi=0.6$ for this particular
$L_z$, but its origin (a three-bounce orbit in the $\rho-z$
plane) as well as its effect are quite different from
the prolate shape. In fact, all downward-diffusing trajectories first cross
$\sinchi_c$ in the vicinity of this island because phase-space flow
(see subsection 6.2) roughly follows a V-shaped curve
connecting the three islands located at $\sinchi=0.825$ ($\theta=\pm
0.5$) and $\sinchi=0.6$ ($\theta=0$). Hence escape is
concentrated in the equatorial region. This remains true even if the
island actually intersects the critical line (as is the case for
higher $L_z$), thus blocking escape
right at $\theta=0$ -- in that case escape still occurs very close to
the island.

Our explanation for the differing intensity profiles is as follows.
First, it is reasonable to assume that all modes for which classical escape is completely
forbidden will lase.  As $L_z$ decreases and classical escape begins to occur there
will exist modes which are lower Q, but still high enough Q to lase.  These modes will
have highly directional emission in the polar angle (they will of course emit uniformly
in the azimuthal angle).  In the prolate case as $L_z$ decreases the first regions of chaos 
which connect the WG orbits to the critical line do so only in small intervals around 
$\theta=30^{\circ}$ and $\theta=150^{\circ}$ (see Figs.~\ref{fig15}(c)
and \ref{fig16}(a)). 
These values of $L_z$ will correspond to the lower Q lasing modes just discussed and will
thus emit in small bright bands around these latitudes.  In contrast, in the oblate droplets
phase space flow curves touch the critical line near the equator (see Fig.~\ref{fig16}(b))
and the lower Q lasing modes will emit around $\theta=0$.
However the total lasing intensity will be due to all the lasing modes
and there will
typically also be many
high $L_z$ modes of higher Q which can emit at the equator through the same processes
as in the undeformed droplet.  But unlike the undeformed case, these modes are competing
with lower Q lasing modes corresponding to lower $L_z$ which therefore have higher gain
and capture more of the pump energy.  Since more of the pump energy goes into these
highly anisotropic modes whose emission pattern differs strongly between the prolate and 
oblate shapes, these shapes show essentially the intensity profiles predicted by the ray-optics
model for the intermediate values of $L_z$ corresponding to slow classical escape and therefore
low but above-threshold Q values.    

To produce a numerical simulation of the cumulative directionality
of all the lasing modes with classical escape, we consider ensembles
of starting conditions homogenously distributed in $\phi$ as in section
6.  If the average $Q$ of
an ensemble starting at some $\sin\chi_0$ is above the lasing threshold,
we record the resulting escape directionality; otherwise we discard that
particular ray bundle. Repeating this for uniformly spaced 
$\sinchi_0\in[\sinchi_c,1]$ and $L_z/\rho(0)\in[0,1]$, one obtains
Fig.~\ref{fig16}(c). It can be seen that maximum emission occurs near the
equator for the oblate shape, and around $30^{\circ}$ away from the
poles in the prolate droplet. The emission peak of the oblate
droplet is split because of the island shown in  Fig.~\ref{fig16}(b),
but this structure is not seen in Fig.~\ref{fig14}. Nevertheless,
the ray theory is clearly able to account for the overall location of
the emission maxima seen experimentally.
The discrepancies in fine structure may be resolvable by modifying the 
parametrization in Eq.~(\ref{shape3d}) which
we chose to model the shape. More experiments are required to check if
the above splitting does indeed occur in droplets at different
oscillation stages.

\section{Summary and Conclusions}
In this chapter we have begun to develop a theory of a new class of asymmetric resonant cavities
(ARCs).   A large class of deformations of dielectric spheres or cylinders were shown to lead to 
resonators with similar properties.  This universality can be understood in terms of partially
chaotic ray dynamics described by KAM theory which we have incorporated into a ray-optics model
for the whispering gallery modes of ARCs.  The Q of these modes decreases sharply at a threshold 
deformation and above this deformation the emission is very directional.  The theory provides
a natural explanation for the anisotropic lasing intensity profile of deformed lasing droplets.
Systematic experimental investigation of ARCs has just begun, and only for the case of lasing
droplets; further experiments are of paramount importance at this point.  
ARCs may provide tests and practical applications for ideas from KAM theory and quantum chaos.
We have proposed a number of potential applications of ARCs which we conjecture may combine the 
advantages of microsphere 
and microdisk cavities with the directional properties of Fabry-Perot cavities.  The feasibility
of these applications cannot be judged unless significant effort is put into the experimental
study of solid ARCs.

\section{Acknowledgements}
We would like to acknowledge the major contribution of Attila Mekis to the section on lasing
droplets and ray directionality, 
as well as valuable discussions with Henrik Bruus and Dima Shepelyansky.
We have also benefited greatly from close contact and collaboration with Richard Chang and
his experimental group, particularly Gang Chen.  This work was partially supported by 
the NSF under grant no. DMR-9215065.


\begin{thebibliography}{111}
\bibitem{WG}
The term whispering gallery mode has been adopted in optics because of
the similarity of these modes to analogous modes in acoustics which
were first understood by Lord Rayleigh.  These modes have
high-energy density only in a small annulus at the perimeter of the
cavity and scatter very little so that in an acoustic cavity (e.g. St.
Paul's cathedral) it is possible to whisper and be heard clearly at
the opposite side of the cavity, but not in its center.  In the
context of the optical study of microdroplets these modes are often
referred to as ``morphology-dependent resonances'' (MDR's).  In this
chapter we shall use the term whispering gallery mode to refer to all
such modes.
\bibitem{ellipse} The only relevant exception is the case of elliptical
deformations (discussed in subsection 3.2) which may be
treated analytically, see [C.~Yeh, {\it J.~Math.~Phys.~}{\bf 4}, 65
(1963)] but which is not typical
because the ray dynamics remains integrable for all deformations.
\bibitem{photoatom}
It is amusing to note an analogy between the microwave ionization of
hydrogen and our current topic.  The dielectric microspheres which
trap a long-lived photon are sometimes referred to as ``photonic
atoms''; thus one could say we are studying the classical
``ionization'' of the photonic atom due to a perturbation of its
shape.
\end{thebibliography}
\end{document}